\documentstyle[11pt]{article}

\begin{document}
\topmargin 0pt
\oddsidemargin 5mm
\setcounter{page}{1}
\begin{titlepage}
\hfill Preprint YERPHI-1552(1)-2000

\vspace{0.5cm}
\begin{center}

{\bf Axial Symmetry and Bound States of Particles with Anomalous Magnetic Moment }\\
\vspace{5mm}
{\large R.A. Alanakyan \\}
\vspace{2mm}
{(C) All Rights Reserved 1999\\}

\vspace{5mm}
{\em Theoretical Physics Department,
Yerevan Physics Institute,
Alikhanian Brothers St.2,
Yerevan 375036, Armenia\\}
 {E-mail: alanak@lx2.yerphi.am\\}

\newpage
\end{center}
\vspace{3mm}
\centerline{{\bf{Abstract}}}

Considered bound states of fermions with an anomalous magnetic
moments in the field of charged cylinder.Also obtained second order
equations for axially symmetric $Z_0(r)$-boson field, radial axially symmetric magnetic
field, obtained second order equations for case of  bound states of fermions
in the radial electric and magnetic fields.
\vspace{3mm}
\vfill
\centerline{{\bf{Yerevan Physics Institute}}}
\centerline{{\bf{Yerevan 2000}}}

\end{titlepage}
                {\bf 1.Introduction}

As known, the dynamics of the neutral  fermions with anomalous magnetic
moments are described  by Dirac equation with non-minimal coupilngs
of neutral fermions with electromagnetic field \cite{LL4},\cite{AB}:
\begin{equation}
\label{A18}
(\hat{k}-m+\mu (\vec{\Sigma}\vec{B}-i\vec{\alpha}\vec{E})+iq(\vec{E}\vec{B})\gamma_5 )\psi(k)=0,
\end{equation}
where  $\mu $-is anomalous magnetic moment,
$\frac{1}{2}\vec{\Sigma}$ is spin operator,and defined by formula (21,21) of \cite{LL4}
operator $\vec{\alpha}$ is defined by formula (21,20) of
\cite{LL4}.The last term described by lagrangian:
\begin{equation}
\label{A4}
L=iq(\vec{E}\vec{H})\bar{\psi} \gamma_5 \psi
\end{equation}

in Dirac equations is obtained in
\cite{DF} and depends only on $r$\cite{RA1} in case of monopole
(for monopoles see e.g. \cite{R} and references therein) which have both electric
 and magnetic fields.

In this article we consider cylindrically symmetric bound states
 and resonances  of particles with anomalous magnetic moments.

It is of interest to consider several special cases in particular:

1)Particle energy levels in pure electric axial field case ($\vec{E}=(E_r(r),0,0)$)
which created e.g. by
homogeneously charged cylinder.

In accordance with \cite{BT} we have:
\begin{equation}
\label{A5}
E_r=\frac{2\sigma}{r}\quad at \quad r>R
\end{equation}
and
\begin{equation}
\label{A5}
E_r=\frac{2\sigma r}{R^2}\quad at \quad r<R
\end{equation}
where $\sigma$ is density of charge of the unit of length of the cylinder.

Also we consider  second order
equations for axially symmetric $Z_0(r)$-boson field, radial axially symmetric magnetic
field obtain second order equations for case of  bound states of fermions
in the radial electric and magnetic fields \cite{RA1}, and also
obtain second order equations for case of pseudoscalars described by last term
in Dirac equation (1).

{\bf Radial axially symmetric electric field}

In ref.\cite{RA2} has been obtained  the system of  equations for radial
 functions (see below  formulas (20)-(23)in which we add also radial magnetic field besides radial
electric field).Below will be presented the second order equations which obtained
 after excluding  two of four radial functions:
\begin{equation}
\label{A30}
(\frac{1}{r}\frac{d}{dr}r\frac{d}{dr}+\epsilon^2-m^2-p_z^2
 -4\pi\rho -( \frac{l}{r}-\mu E)^2)f_1(r)+2i\mu Ep_zf_2(r)=0
\end{equation}
\begin{equation}
\label{A30}
(\frac{1}{r}\frac{d}{dr}r\frac{d}{dr}+\epsilon^2-m^2-p_z^2
 -4\pi\rho -( \frac{(l+1)}{r}+\mu E)^2)f_2(r)-2i\mu Ep_zf_1(r)=0
\end{equation}

The presence of the $i$ mean that one of the radial functions must be purely imagine.

Analogous system of equations obtained if we esclude $\phi$ and
consider $\chi$ as $\chi=(e^{il \phi}f_3(r),e^{i(l+1)
\phi}f_4(r))$:

\begin{equation}
\label{A30}
(\frac{1}{r}\frac{d}{dr}r\frac{d}{dr}+\epsilon^2-m^2-p_z^2
 +4\pi\rho -( \frac{l}{r}-\mu E)^2)f_3(r)+2i\mu Ep_zf_4(r)=0
\end{equation}
\begin{equation}
\label{A30}
(\frac{1}{r}\frac{d}{dr}r\frac{d}{dr}+\epsilon^2-m^2-p_z^2
 +4\pi\rho -( \frac{(l+1)}{r}+\mu E)^2)f_4(r)-2i\mu Ep_zf_3(r)=0
\end{equation}

In case of electric field created by charged line or charged
cylinder at $r>a$ we obtain:
\begin{equation}
\label{A30}
(\frac{1}{r}\frac{d}{dr}r\frac{d}{dr}+\epsilon^2-m^2-p_z^2
 - \frac{(l-2\mu\sigma)^2}{r^2})f_1(r)+ \frac{4i\mu \sigma p_z}{r}f_2(r)=0
\end{equation}
\begin{equation}
\label{A30}
(\frac{1}{r}\frac{d}{dr}r\frac{d}{dr}+\epsilon^2-m^2-p_z^2
  - \frac{(l+1+2\mu\sigma)^2}{r^2})f_2(r)-\frac{4i\mu \sigma p_z}{r}f_1(r)=0
\end{equation}
Inside homogeneously charged cylinder we obtain:
\begin{equation}
\label{A30}
(\Omega
 - (\frac{(l)^2}{r^2}+\frac{4 \mu^2 \sigma^2 r^2}{a^4}+\frac{4 \mu \sigma l}{a^2}))f_1(r)
 + \frac{4i \mu \sigma p_zr}{a^2}f_2(r)=0
\end{equation}
\begin{equation}
\label{A30}
(\Omega-(\frac{(l+1)^2}{r^2}+\frac{4 \mu^2 \sigma^2 r^2}{a^4}
+\frac{4 \mu \sigma (l+1)}{a^2}))f_2(r)-\frac{4i\mu \sigma p_z}{r}f_1(r)=0
\end{equation}
where $\Omega=\frac{1}{r}\frac{d}{dr}r\frac{d}{dr}+\epsilon^2-m^2-p_z^2$
we see that at $p_z=0$ both equations decouples and every of them
is the same as for 2-dimensional harmonic oscillator
and in accordance with  \cite{LL3} we have the following energy levels:
\begin{equation}
\label{A30}
\epsilon^2=m^2+\frac{4 \mu \sigma }{a^2}+\frac{8 |\mu \sigma| }{a^2}(n_r+\frac{l+|l|+1}{2})
\end{equation}
Analogously for second equation in result for energy levels we must replace:

$l \rightarrow (l+1)$

The particle is localized at distances $r_H \sim
(\frac{4\mu\sigma}{a^2})^{-\frac{1}{2}}$ or smaller.
Of course our consideration is available only if size of the
particle localization is essentially smaller than the radius of
the cylinder($r_H<<a$).The  wave functions at large distances are
suppressed by exponent $e^{-\frac{r^2}{r_H^2}}$.

Above was considered energy levels of neutrons in case infinite radius of the cylinder.
It is of interest to consider also case of the finite radius of the cylinder.

For this purpose we must find wave function inside and outside of the cylinder
and in this case energy levels of neutrons will be defined from condition:

\begin{equation}
\label{A30}
\frac{R'_{r>a}(a) }{R_{r>a}(a)}=\frac{R'_{r<a}(a) }{R_{r<a}(a)}.
\end{equation}

Inside cylinder as we seen above we obtain equation which is equivalent to the
 2-dimensional harmonic oscillator  and radial wave functions are expresses through
degenerate hypergeometric function:
\begin{equation}
\label{A30}
R_{r<a}(r)=C_1x^{\frac{|l|}{2}}e^{-\frac{x}{2}}
F(\frac{l+|l|+1}{2}-\frac{\epsilon^2-m^2-4\pi\mu\rho}{2m \omega},|l|+1,x)
\end{equation}
where $\omega=\frac{4|\mu \sigma|}{m a^2}$,$x=m \omega r^2$.
Outside of the cylinder radial wave functions are expresses through
Macdonald's function:

\begin{equation}
\label{A30}
R_{r<a}(r)=C_2K_{|l+2\mu\sigma|}(\sqrt{|\epsilon^2-m^2|}r)
\end{equation}

Below we consider case of the charged line.

In the limit:
\begin{equation}
\label{A30}
\mu \sigma <<1
\end{equation}

we obtain Coulomb-like spectrum for energy levels of neutrons in the field of charged line.

 Indeed, if $l$ is not equal to the
$0,-1$ we can put in equations (5), (6) $l-2\mu\sigma \approx l$ in
the limit $\mu \sigma <<1$.At $l=0,-1$ must be
\begin{equation}
\label{A30}
\frac{\mu^2 \sigma^2}{r_B^2} << \frac{\mu \sigma p_z}{r_B}
\end{equation}
where $r_B=\frac{1}{\mu \sigma p_z}$ is Bohr radius (see
below).Substituting $r_B$ in (18) we obtain again the condition $\mu \sigma
<<1$.Thus, in the limit $\mu \sigma <<1$  at $l=0,-1$  term $=(0+\mu \sigma)^2r^{-2}$
must be neglected, and  we obtain for all $l$ the following equations:
\begin{equation}
\label{A30}
(-\frac{1}{r}\frac{d}{dr}r\frac{d}{dr}+\epsilon^2-m^2-p_z^2
 + \frac{l^2}{r^2})f_1(r)+ \frac{4\mu \sigma p_z}{r}if_2(r)=0
\end{equation}
\begin{equation}
\label{A30}
(-\frac{1}{r}\frac{d}{dr}r\frac{d}{dr}+\epsilon^2-m^2-p_z^2
  + \frac{(l+1)^2}{r^2})if_2(r)+\frac{4\mu \sigma p_z}{r}f_1(r)=0
\end{equation}
This equations are similar to equations obtained in \cite{P} where
has been consider non-relativistic neutrons energy levels in the
magnetic field $\vec{H}=(0,H_{\phi}(r)=2I/r,0)$
if we making the following replacement:
\begin{equation}
\label{A5}
I \rightarrow  \frac{\sigma p_z}{m}
\end{equation}

Thus, energy levels is also same and in accordance with result of \cite{P}
and substitution (21) are defines by quantum number $n$:
\begin{equation}
\label{A30}
\epsilon^2_n=m^2+p_z^2-\frac{ \mu^2 \sigma^2 p_z^2  }{n^2}
\end{equation}
It must be noted, our consideration is
relativistic,only assumption $\mu \sigma <<1$ has been used.
In the near future we will present the solution where $\mu \sigma$
is not small.

{\bf Radial Axially Symmetric Magnetic +Radial Axially Symmetric electric field and
bound states of neutral fermions with an anomalous magnetic moments}

In this paper we consider also  bound states of neutral fermions
with an anomalous magnetic moments in radial axially symmetric magnetic
field $\vec{H}=(\frac{x}{\sqrt{x^2+y^2}}H(r),\frac{y}{\sqrt{x^2+y^2}}H(r),0)$
which analogously to the above considered electric field in case of cylinder
 has the following form:
\begin{equation}
\label{A5}
H_r(r)=\frac{2\sigma_m}{r}\quad at \quad r>R
\end{equation}
and
\begin{equation}
\label{A5}
H_r(r)=\frac{2\sigma_m r}{R^2}\quad at \quad r<R
\end{equation}
where $\sigma_m$ is density of magnetic charge of the unit of length of the cylinder.
\begin{equation}
\label{A5}
(\epsilon-m)f_1+\mu H f_2-p_3 f_3-i(\frac{d}{dr}+\frac{l}{r}-\mu E)f_4=0
\end{equation}
\begin{equation}
\label{A5}
\mu H f_1+(\epsilon-m)f_2-i(\frac{d}{dr}-\frac{l-1}{r}-\mu E)f_3+p_3 f_4=0
\end{equation}
\begin{equation}
\label{A5}
p_3 f_1+(\frac{d}{dr}+ \frac{l}{r}+\mu E)f_2-(\epsilon+m) f_3+\mu H f_4=0
\end{equation}
\begin{equation}
\label{A5}
-i(\frac{d}{dr}- \frac{l-1}{r}+\mu E)f_1- p_3 f_2+\mu H f_3-(\epsilon+m)f_4=0
\end{equation}

It is interesting to notice that equations for radial magnetic
field is similar to equations for magnetic field
$\vec{H}=(0,H_{\phi}(r),0)$ considered in \cite{RA2}.

In non-relativistic approximation we have the following system of
equations (instead Dirac equation has been used Pauli equation
with non-relativistic spinor
$\phi=(f_1(r)e^{i(l-1)}\phi,f_2(r)e^{il})$):
\begin{equation}
\label{A5}
\frac{1}{2m}\Omega_1(l-1)f_1(r)+\mu H(r)f_2(r)=0
\end{equation}
\begin{equation}
\label{A5}
\frac{1}{2m}\Omega_1(l)f_2(r)+\mu H(r)f_1(r)=0
\end{equation}

This equations are similar to equations obtained in \cite{P} where
has been consider non-relativistic neutrons energy levels in the
magnetic field $\vec{H}=(0,H_{\phi}(r)=2I/r,0)$.
It is seen from the following replacement in above derived equations (),():
\begin{equation}
\label{A5}
f_2\rightarrow -if_2, \sigma_m \rightarrow I
\end{equation}

Thus, energy levels is also same and in accordance with result of \cite{P}
are defines by quantum number $n$:
\begin{equation}
\label{A30}
E_n=-\frac{(\mu \sigma_m)^2m }{2n^2}
\end{equation}

{\bf Fermions with  anomalous magnetic moments in magnetic field $\vec{H}=(0,0,H_z(r)=H(r))$ }

In component form
($\psi^T=(f_1(r)e^{i(l-1)\phi},f_2(r)e^{i(l)\phi},f_3(r)e^{i(l-1)\phi},f_4(r)e^{i(l)\phi})$)
 the equations has been obtained in \cite{RA1}:
\begin{equation}
\label{A5}
(\epsilon-m+\mu H)f_1-p_3f_3-p_-f_4=0
\end{equation}
\begin{equation}
\label{A5}
(\epsilon-m-\mu H) f_2-p_+f_3+p_3 f_4=0
\end{equation}
\begin{equation}
\label{A5}
p_3 f_1+p_-f_2+(-\epsilon-m+\mu H) f_3=0
\end{equation}
\begin{equation}
\label{A5}
p_+f_1- p_3 f_2+(-\epsilon-m-\mu H) f_4=0
\end{equation}
where

\begin{equation}
\label{A5}
p_-=-i(\frac{d}{dr}+\frac{l}{r})
\end{equation}

\begin{equation}
\label{A5}
p_+=-i(\frac{d}{dr}-\frac{l-1}{r})
\end{equation}
Excluding two of four components we obtain the system of two second order equations:
\begin{equation}
\label{A5}
i(2\mu H(r)(\frac{d}{dr}-\frac{l-1}{r})+4\pi \mu j_{\phi}(r))f_1(r)
+((m+\mu H(r))^2+\Omega_1(l))f_4(r)=0
\end{equation}

\begin{equation}
\label{A5}
i(2\mu H(r)(\frac{d}{dr}+\frac{l}{r})+4\pi\mu j_{\phi}(r))f_4(r)
+((m-\mu H(r))^2+\Omega_1(l-1))f_1(r)=0
\end{equation}
where $\Omega_1(l)=-\frac{1}{r}\frac{d}{dr}r\frac{d}{dr}+\frac{l^2}{r}+p^2_z+m^2-\epsilon^2$,
 $\vec{j}=(0,j_\phi(r),0)$ is density of current.

{\bf Majorana neutrinos}

In this paper we consider second order  equations for Majorana
neutrinos($g_V=0$).For this purpose it is necessary writing Dirac equation:
\begin{equation}
\label{A18}
(\hat{k}-g_A\hat{Z}\gamma_5-m)\psi(k)=0,
\end{equation}
in component form and excluding one of the component.

The final result for equations of the second order is following:
\begin{equation}
\label{A5}
T(\kappa)f_1+g_A(Z_0'(r)f_2+2Z_0(r)(f_2'(r)+\frac{1-\kappa}{r}f_2(r))=0
\end{equation}

\begin{equation}
\label{A5}
T(\kappa-1)f_2-2g_AZ_0(r)(f_1'(r)+\frac{1+\kappa}{r}f_1(r))=0
\end{equation}
where
\begin{equation}
\label{A5}
T(\kappa)= \epsilon^2+\frac{1}{r^2}\frac{d}{dr}r^2\frac{d}{dr}-\frac{\kappa(\kappa+1)}{r^2}
-g^2_AZ_0^2(r)-m^2,
\end{equation}

\begin{equation}
\label{A23}
\kappa=l(l+1)-j(j+1)-\frac{1}{4},
\end{equation}

We see that term $g^2_AZ_0^2(r)$ is always repulsive.
In cylindrical symmetry case choosing two-component spinor $\phi$
 as $\phi=(f_1(r)e^{il}\phi,f_2(r)e^{i(l+1)})$) we obtain
 the following system of the two  second order  equations:
\begin{equation}
\label{A5}
(P(l)-2g_AZ_0(r)p_z)f_1+g_A(2Z_0(r)(\frac{d}{dr}+\frac{1+l}{r})+Z_0'(r))if_2(r)=0
\end{equation}
\begin{equation}
\label{A5}
(P(l+1)+2g_AZ_0(r)p_z)if_2-g_A(2Z_0(r)(\frac{d}{dr}-\frac{1}{r})+Z_0'(r))f_1(r)=0
\end{equation}
where
$P(l)= \epsilon^2+\frac{1}{r}\frac{d}{dr}r\frac{d}{dr}-\frac{l^2}{r^2}-p^2_z-g^2_AZ_0^2-m^2$.

As known, anomalous magnetic moments of the Majorana neutrino is equal to the zero.
Thus , in  case of Majorana neutrino  the attraction is possible only via interaction (2)
 in radial electric and magnetic fields of monopole (see \cite{RA1} for detailes).
The second order equations in this case (i.e. if only interaction (2) presented)
are following:
\begin{equation}
\label{A5}
(\epsilon^2+\frac{1}{r^2}\frac{d}{dr}r^2\frac{d}{dr}-\frac{\kappa(\kappa+1)}{r^2}
-A^2-m^2)f_1+Aif_2(r)=0
\end{equation}

\begin{equation}
\label{A5}
(\epsilon^2+\frac{1}{r^2}\frac{d}{dr}r^2\frac{d}{dr}-\frac{\kappa(\kappa-1)}{r^2}
-A^2-m^2)if_2+Af_1(r)=0,
\end{equation}
where $A=iqE(r)H(r)$.We see that term $A^2(r)\sim r^{-8}$ is always repulsive, dominate at small
distances and prevent fall down on the center.

In cylindrical symmetry case we have:
\begin{equation}
\label{A5}
(\epsilon^2+\frac{1}{r}\frac{d}{dr}r\frac{d}{dr}-\frac{l^2}{r^2}-p_z^2
-A^2-m^2)f_1+Aif_2(r)=0,
\end{equation}

\begin{equation}
\label{A5}
(\epsilon^2+\frac{1}{r}\frac{d}{dr}r\frac{d}{dr}-\frac{(l+1)^2}{r^2}-p_z^2
-A^2-m^2)f_1+Aif_2(r)=0,
\end{equation}

{\bf Bound states of fermions with  anomalous magnetic moments in radial electric
and magnetic field: second order equations }

Although it is possible to exclude to of four radial functions  from system of
equations of the first order (see  (12)-(16) in \cite{RA1})
 which defines energy levels of the neutral fermion with anomalous magnetic moment
, much more convenient  in order to obtain the second order equations
to start from  Dirac equation in component form:
\begin{equation}
\label{A5}
(\epsilon-m+\mu \vec{\sigma}\vec{H})\phi-\vec{\sigma}(\vec{p}+i\mu \vec{E})\chi=0
\end{equation}

\begin{equation}
\label{A5}
(-\epsilon-m+\mu \vec{\sigma}\vec{H})\chi+\vec{\sigma}(\vec{p}-i\mu \vec{E})\phi=0
\end{equation}

Excluding e.g. component $\chi$  and presenting $\phi$ as linear
\newpage
combinations of spherical spinors with different $P$-parity
(analogously \cite{RA3}, because in studied case P-parity violation take place ) \footnote{it must be stressed that
besides $P$-parity violation also presented $T$-parity violation
(term $\vec{\Sigma} \vec{r}H$ in Dirac equation is $P$- and $T$ -parity violating)
 and purely imagine character of the two of four
radial function in equations (12)-(16) of the \cite{RA1}is connected with this circumstances.
Also, in (55),(56) e.g. $f_2$ must be purely imagine, $f_1$ must be real. }:
\begin{equation}
\label{A20}
\phi^T=f_1(r)\Omega_{jlM}(\vec{n})+(-1)^{\frac{1+l-l'}{2}}f_2(r)\Omega_{jl'M}(\vec{n})
\end{equation}
we obtain the following system  of the second order equations:
\begin{equation}
\label{A5}
2m \mu Hif_2-T_+(\kappa)f_1+Q(\kappa)(a_+S_+(-\kappa)if_2-b_+S_+(\kappa)f_1)
\end{equation}
\begin{equation}
\label{A5}
2m \mu Hf_1-T_+(-\kappa)if_2+Q(-\kappa)(a_+S_+(\kappa)f_1-b_+S_+(-\kappa)if_2)
\end{equation}

Analogously, excluding $\phi$ and presenting $\chi$  as:

\begin{equation}
\label{A5}
\chi=g_1(r)\Omega_{jlM}(\vec{n})+(-1)^{\frac{1+l-l'}{2}}g_2(r)\Omega_{jl'M}(\vec{n}))
\end{equation}
we obtain the system for $g_{1,2}$:
\begin{equation}
\label{A5}
2m \mu Hig_2-T_-(\kappa)g_1-Q(\kappa)(a_-S_-(-\kappa)ig_2-b_-S_-(\kappa)g_1)
\end{equation}
\begin{equation}
\label{A5}
2m \mu Hg_1-T_-(-\kappa)ig_2-Q(-\kappa)(a_-S_-(\kappa)g_1-b_-S_-(-\kappa)ig_2)
\end{equation}

where:

\begin{equation}
\label{A5}
Q(\kappa)=4\pi \mu \rho_m-\frac {2\mu H(r)}{r}(1\pm \kappa),
\end{equation}

\begin{equation}
\label{A5}
S_{\pm}(\kappa)= \frac{d}{dr}+ \frac{1\pm \kappa}{r}\pm \mu E,
\end{equation}

\begin{equation}
\label{A5}
T_{\pm}(\kappa)=\epsilon^2-m^2+
\frac{1}{r^2}\frac{d}{dr}r^2\frac{d}{dr}-\frac{\kappa(\kappa+1)}{r^2}-\mu^2(E^2+H^2)
\pm 4\mu \pi \rho \mp \frac{2 \mu E}{r}(1\pm \kappa)
\end{equation}
\begin{equation}
\label{A5}
a_{\pm}=\frac{\epsilon \pm m}{(\epsilon \pm m)^2-\mu^2 H^2}
\end{equation}

\begin{equation}
\label{A5}
b_{\pm}=\frac{\pm \mu H }{(\epsilon \pm m)^2-\mu^2 H^2}
\end{equation}

 During derivation of this formulas we take into account that
$div \vec{E}=4 \pi \rho$,$div \vec{H}=4 \pi \rho_m$.

In nonrelativistic limit terms which are proportional to the $a_{\pm},b_{\pm}$
are small and may be neglected.

We see, that terms $(\mu E(r))^2\sim r^{-4}$
 $(\mu H(r))^2\sim r^{-4}$ are repulsive and prevent fall down on
 the center.

The author express his sincere gratitude to Zh.K.Manucharyan and
E.B.Prokhorenko  for helpful discussions.

\end{document}